# Massively Parallel Sort-Merge Joins in Main Memory Multi-Core Database Systems

Martina-Cezara Albutiu Alfons Kemper Thomas Neumann

Technische Universität München
Boltzmannstr. 3
85748 Garching, Germany
firstname.lastname@in.tum.de

## ABSTRACT

Two emerging hardware trends will dominate the database system technology in the near future: increasing main memory capacities of several TB per server and massively parallel multi-core processing. Many algorithmic and control techniques in current database technology were devised for disk-based systems where I/O dominated the performance. In this work we take a new look at the well-known sort-merge join which, so far, has not been in the focus of research in scalable massively parallel multi-core data processing as it was deemed inferior to hash joins. We devise a suite of new massively parallel sort-merge (MPSM) join algorithms that are based on partial partition-based sorting. Contrary to classical sort-merge joins, our MPSM algorithms do not rely on a hard to parallelize final merge step to create one complete sort order. Rather they work on the independently created runs in parallel. This way our MPSM algorithms are NUMA-affine as all the sorting is carried out on local memory partitions. An extensive experimental evaluation on a modern 32-core machine with one TB of main memory proves the competitive performance of MPSM on large main memory databases with billions of objects. It scales (almost) linearly in the number of employed cores and clearly outperforms competing hash join proposals – in particular it outperforms the "cutting-edge" Vectorwise parallel query engine by a factor of four.

## 1. INTRODUCTION

Increasing main memory capacities of up to several TB per server and highly parallel processing exploiting multi-core architectures dominate today's hardware environments and will shape database system technology in the near future. New database software has to be carefully targeted against the upcoming hardware developments. This is particularly true for main memory database systems that try to exploit the two main trends – increasing RAM capacity and core numbers. So far, main memory database systems were either designed for transaction processing applications, e.g., VoltDB [25], or for pure OLAP query processing [4]. However, the upcoming requirements for so-called real-time or operational business intelligence demand complex query processing in "real time" on main memory resident data. SAP's Hana [10] and our hybrid OLTP&OLAP database system HyPer [16], for which the presented massively parallel join algorithms were developed, are two such databases. The query processing of in-memory DBMSs is no longer I/O bound and, therefore, it makes sense to investigate massive intra-operator parallelism in order to exploit the multi-core hardware effectively. Only query engines relying on intra-query and intra-operator parallelism will be able to meet the instantaneous response time expectations of operational business intelligence users if large main memory databases are to be explored. Single-threaded query execution is not promising to meet the high expectations of these database users as the hardware developers are no longer concerned with speeding up individual CPUs but rather concentrate on multi-core parallelization.

Consequently, in this paper we develop a new sort-based parallel join method that scales (almost) linearly with the number of cores. Thereby, on modern multi-core servers our sort-based join outperforms hash-based parallel join algorithms which formed the basis for multi-core optimization in recent proposals. The well-known radix join algorithm of MonetDB [19] pioneered the new focus on *cache locality* by repeatedly partitioning the arguments into ever smaller partitions. The recursive sub-partitioning, rather than directly partitioning into small fragments, preserves TLB cache locality by restricting the random write of the partitioning phase to a small number of pages whose addresses fit into the TLB cache. The join is carried out on small cache-sized fragments of the build input in order to avoid cache misses during the probe phase. Because of this cache-affine behavior the radix join became the basis for most work on multi-core parallel join implementations, e.g., [17, 14]. In addition to the cache locality, He et al. [14] and Kim et al. [17] also focussed on low synchronization overhead and avoidance of dynamic memory allocation. Both aspects were achieved by computing histograms of the data to be partitioned and then deriving the prefix sums to determine the exact array positions into which parallel threads write their partitioned data. Unfortunately, merely relying on straightforward partitioning techniques to maintain cache locality and to keep all cores busy will not suffice for the modern hardware that scales main memory via non-uniform memory access (NUMA). Besides the multi-core parallelization also the RAM and cache hierarchies have to be taken into account. In particular the NUMA division of the RAM has to be considered carefully. The whole NUMA system logically divides into multiple nodes, which can access both local and remote memory resources. However, a processor can access its own local memory faster than non-local memory, that is, memory local to another processor or memory shared between processors. The key to scalable, high performance is **data placement** and **data movement** such that threads/cores work mostly on local data – called NUMA-friendly data processing.

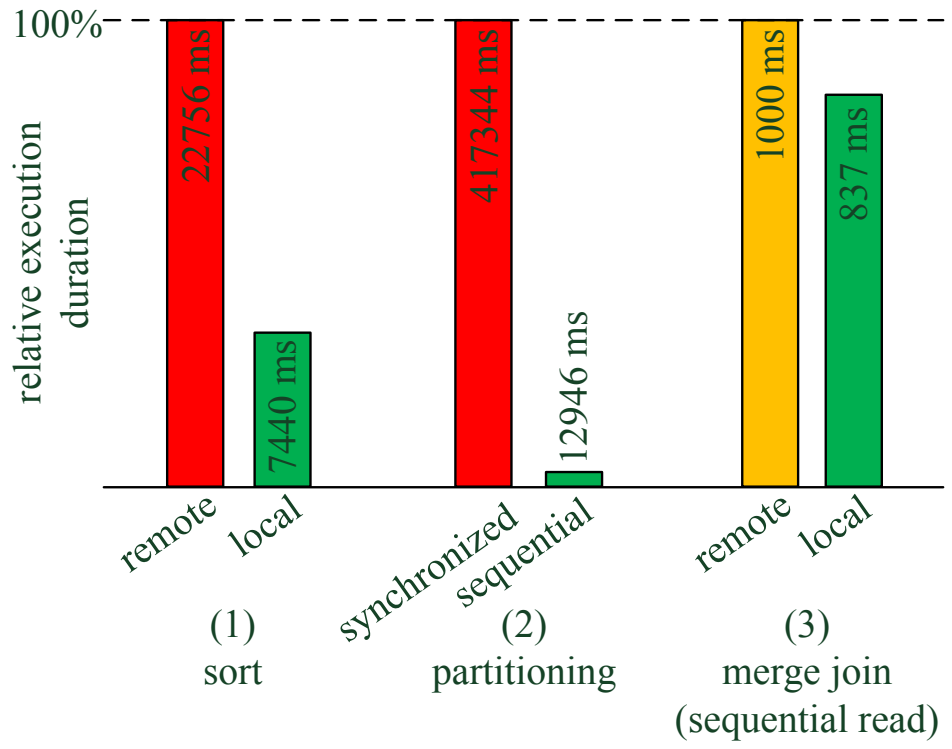

Figure 1: Impact of NUMA-affine versus NUMA-agnostic data processing

To back up this claim, Figure 1 shows the results of a few micro-benchmarks we ran on a one TB main memory machine with 32 cores (cf. Section 5, Figure 11). We therefore instantiated 32 threads to work on one relation with a total of $1600M$ (throughout the paper we use $M = 2^{20}$) tuples, each consisting of a 64-bit sort key and a 64-bit payload, in parallel. (1) We first chunked the relation and sorted the chunks of $50M$ tuples each as runs in parallel. In the "green" NUMA-affine benchmark, the sorting of each core was performed in the local NUMA RAM partition whereas in the unfavorable "red" case the sort was performed on a globally allocated array. We observe a severe performance penalty of a factor of three if NUMA boundaries are ignored. (2) We then analyzed the performance penalty of fine-grained synchronization. For this the 32 threads partitioned the global relation into 32 chunks each being stored as an array. In the "red" experiment the next write position was individually read from a (test-and-set) synchronized index variable of the corresponding partition array. In the "green" experiment all threads were allocated precomputed sub-partitions that could be written sequentially without synchronization. This experiment proves that fine-grained synchronization (even with wait-free test-and-set variables) is a "no-go" for scalable data processing. (3) Finally, in the last microbenchmark we analyzed the tolerable performance penalty of sequentially scanning remote memory in comparison to local memory. Each of the 32 parallel threads merge joins two chunks of $50M$ tuples each. Thereby, each thread works on one local run. The second run is either in remote ("yellow") or local ("green") NUMA partitions. The negative impact of the second chunk being accessed remotely compared to the second chunk being local, too, is mitigated by the hardware prefetcher as the accesses are sequential. We thus conclude that sequential scans of remote memory are acceptable from a performance perspective.

This observation and further micro-benchmarks led us to state the following three rather simple and obvious rules (called "commandments") for NUMA-affine scalable multi-core parallelization:

- C1 *Thou shalt not write thy neighbor's memory randomly* – chunk the data, redistribute, and then sort/work on your data locally.
- C2 *Thou shalt read thy neighbor's memory only sequentially* – let the prefetcher hide the remote access latency.
- C3 *Thou shalt not wait for thy neighbors* – don't use fine-grained latching or locking and avoid synchronization points of parallel threads.

By design, the massively parallel sort-merge join algorithms (called MPSM) we present in this paper obey all three commandments whereas the previously proposed hash join variants violate at least one of the commandments and, therefore, exhibit scalability problems of various forms.

We will show that the carefully engineered NUMA-friendly MPSM exhibits an outstanding performance when compared to the Wisconsin hash join [1] and Vectorwise [15]. Our performance evaluation proves the scalability of MPSM for very large main memory databases with hundreds of GB data volume. For large numbers of cores (up to 32) MPSM outperforms the recently proposed hash-based Wisconsin join by up to an order of magnitude. MPSM scales (almost) linearly in the number of cores and compared to the TPC-H endorsed "world champion" query processor Vectorwise even achieves a factor of four.

The remainder of the paper is structured as follows: In Section 2 we depict the basic idea of MPSM in comparison to the radix join and the Wisconsin hash join. In Section 3 we address the concrete implementations of the MPSM concept in detail and in Section 4 we discuss the skew resilience of MPSM. We evaluate MPSM in Section 5 and cover related work in Section 6. Finally, we conclude our work in Section 7.

## 2. THE BASIC IDEA OF MPSM

We will first present the very basic idea of the NUMA-affine MPSM in comparison to radix join and Wisconsin hash join. Later, we will discuss important refinements regarding performance improvement and skew resilience.

The recently proposed Wisconsin hash join [2] is based on a global shared hash table which has to be built across the NUMA partitions by a large number of threads. These concurrent accesses to a single hash table need synchronization via latches. Therefore, during the parallel build phase "commandments" C2 and C3 are violated. During the probe phase random reads to the hash table are performed across the NUMA memory partitions, which again violates C2 as the hardware prefetcher cannot hide the access latency. In Figure 2a we illustrate the random writes and reads within the NUMA partitions using different-colored arrows and the required synchronization with locks.

The radix join of MonetDB [19] and Oracle/Intel [17] writes across NUMA partitions during the initial partitioning phase as illustrated in Figure 2b. The radix join repeatedly partitions the arguments in order to achieve cache locality of the hash table probes despite their random nature.

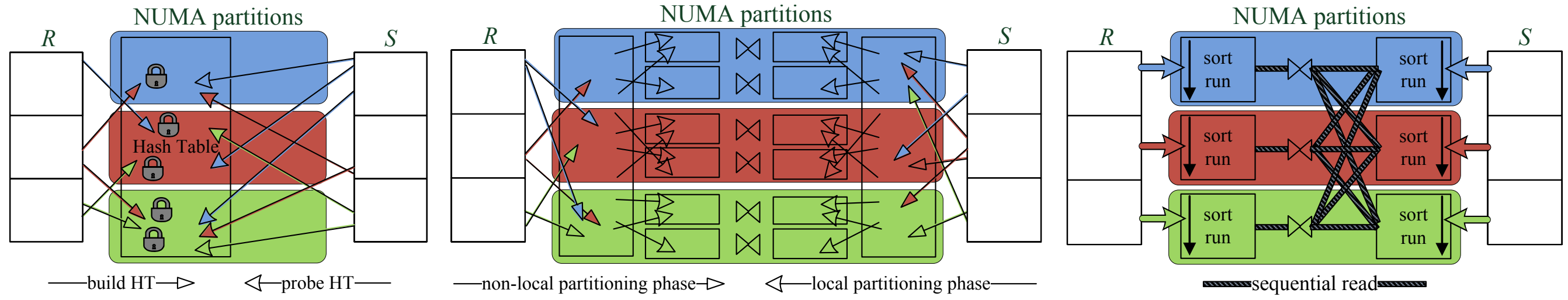

(a) Wisconsin hash join processing (b) Radix join processing (c) MPSM join processing

Figure 2: Comparison of basic join processing of Wisconsin hash join, radix join, and MPSM

Unfortunately, the price for this locality is the partitioning of both join arguments across the NUMA memory during the first partitioning step.

Our massively parallel sort-merge (MPSM) join is designed to take NUMA architectures into account which were not yet in the focus of prior work on parallel join processing for main memory systems. Though, we emphasize that MPSM is oblivious to specific NUMA architectures as it only assumes the locality of a RAM partition for a single core – without relying on multiple cores sharing the locality of RAM partitions or caches. As illustrated in Figure 2c each data chunk is processed, i.e., sorted, locally. Unlike traditional sort-merge joins we refrain from merging the sorted runs to obtain a global sort order and rather join them all in a brute-force but highly parallel manner. We opt to invest more into scanning in order to avoid the hard to parallelize merge phase. Obviously, this decision does not result in a globally sorted join output but exhibits a partial sort order that allows for sort order based subsequent operations, e.g, early aggregation. During the subsequent join phase, data accesses across NUMA partitions are sequential, so that the prefetcher mostly hides the access overhead. We do not employ shared data structures so that no expensive synchronization is required. Therefore, MPSM obeys all three NUMA-commandments by design.

## 2.1 The B-MPSM Algorithm

The basic MPSM algorithm starts by generating sorted runs in parallel. These runs are not merged as doing so would heavily reduce the "parallelization power" of modern multi-core machines. Instead, the sorted runs are simply joined in parallel. In the following, we first describe the MPSM algorithm in its basic form (B-MPSM) which is absolutely insensitive to any kind of skew. It bears some similarity to fragment and replicate distributed join algorithms. However, it only replicates merge join scans of the threads/cores but does not duplicate any data. Then we present an improved MPSM version based on range partitioning of the input by join keys (P-MPSM). Further, the MPSM can effectively be adapted to non-main memory scenarios, i.e., scenarios in which intermediate data must be written to disk. We call this the disk-enabled MPSM algorithm (D-MPSM).

The B-MPSM algorithm is sketched in Figure 3 for a scenario with four worker threads. The input data is chunked into equally sized chunks among the workers, so that for instance worker $W_1$ is assigned a chunk $R_1$ of input $R$ and another chunk $S_1$ of input $S$. Each worker sorts its data chunks, thereby generating sorted runs of the input data in parallel. In the following, we call $R$ the private input and $S$ the public input. After the sorting phase is finished each worker processes only its own chunk of the private input but sequentially scans the complete public input. We will later devise the range partitioned variant where this complete scanning is avoided to speed up the join phase even more beyond parallelization. During run generation (phase 1 and phase 2), each worker thread handles an equal share of both the public and the private input. These phases do not require any synchronization between the workers and are performed in local memory which we have shown to be advantageous for the sort operator (cf. Figure 1). Even if data has to be copied from remote to local chunks this can be amortized by carrying out the first partitioning step of sorting while copying. In phase 3, each worker joins its sorted private input run with the sorted public input runs using merge join. The join phase requires reading non-local memory, however, only sequentially. As we have shown before, sequential scans heavily profit from (implicit processor) prefetching and cache locality and therefore do not affect performance significantly.

The B-MPSM algorithm is absolutely skew resistant and obeys the three "commandments" for NUMA-affine design we stated above: During the run generation phases for public and private input, only local memory is written. In the join phase, all runs (local and remote) are scanned sequentially. Furthermore, B-MPSM requires only one synchronization point as we need to make sure that the public input runs $S_i$ are ready before we start the join phase. Note that the sort phase of the private data $R$ need not be finished before other threads start their join phase. Thus, the synchronization is limited to ensure that all other workers have finished their sorting of the public input chunk before phase 3 (join) is entered. The fact that the output of each worker is a sorted run may be leveraged by subsequent operators like sort-based aggregation. Also, presorted relations can obviously be exploited to omit one or both sorting phases.

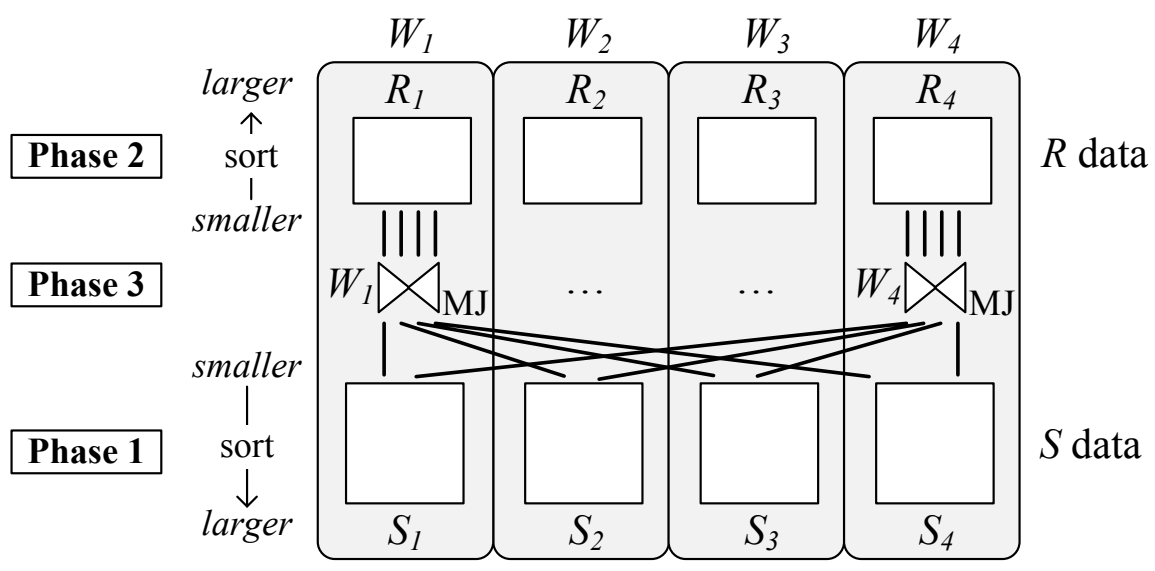

Figure 3: B-MPSM join with four workers $W_i$

## 2.2 Complexity of B-MPSM

B-MPSM basically executes $T$ sort-merge joins in parallel, where $T$ is the number of worker threads. In each of these sort-merge joins, $1/T^{\text{th}}$ of the input relations is processed. A crude complexity approximation per worker $W_i$ results in:

$$\begin{array}{rrl} & {|S|}/{T} \cdot log({|S|}/{T}) & \text{sort chunk } S_i \text{ of size } {|S|}/{T} \\ + & {|R|}/{T} \cdot log({|R|}/{T}) & \text{sort chunk } R_i \text{ of size } {|R|}/{T} \\ + & T \cdot {|R|}/{T} & \text{process run } R_i \text{ for all } S \text{ runs} \\ + & T \cdot {|S|}/{T} & \text{process all } S \text{ runs} \end{array}$$

$$= {|S|}/{T} \cdot log({|S|}/{T}) + {|R|}/{T} \cdot log({|R|}/{T}) + |R| + |S|$$

On the bottom line, each thread sorts "its" chunks of $R$ and $S$ and processes all sorted $S$ runs. Thereby, the own $R$ run is read several ($T$) times as each of the $S$ runs possibly joins with the local run.

The formula above reveals that the sort phases of B-MPSM scale well with the number of worker threads $T$. The join phase, however, requires each worker to process the complete public input regardless of the processing parallelism given. For I/O bound disk-based processing in D-MPSM this is hidden by the I/O latency. However, for pure in-core processing we address this issue by including a prologue phase to range partition and assign the private input data to the workers in a way that allows saving much of the work during the join phase. This variant is called (partitioned) P-MPSM and is explained in detail in this paper.

## 2.3 Sorting

Efficient sorting is decisive for the performance of MPSM. As we deal with (realistic) large join keys and payloads that need to be sorted we cannot utilize the specialized bitonic sorting routines that exploit the SIMD registers [6], as these are limited to 32-bit data types. Instead we developed our own three-phase sorting algorithm that operates as follows:

1. **in-place** Radix sort [18] that generates $2^8 = 256$ partitions according to the 8 most significant bits (MSD). This works by computing a 256 bucket histogram and determining the boundaries of each partition. Then the data elements are swapped into their partition.
2. IntroSort (Introspection Sort) [20]
   - 2.1 Use Quicksort to at most $2 \cdot log(N)$ recursion levels. If this does not suffice, resort to heapsort.
   - 2.2 As soon as a quicksort partition contains less than 16 elements stop and leave it to a final insertion sort pass to obtain the total ordering.

We analyzed that this sorting routine is about 30% faster than, for example, the STL sort method – even when up to 32 workers sort their local runs in parallel. Note that we do not employ synchronization-heavy parallel sorting – each worker sorts a separate chunk of data into a run.

# 3. THE RANGE-PARTITIONED MPSM AND A DISK-BASED VARIANT

So far, we presented the basic concept of MPSM which (in object-oriented terminology) is only an abstract class for several algorithmic specializations. We present two derived implementations:

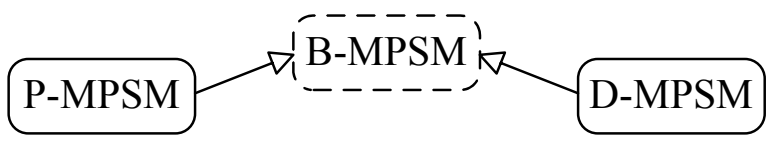

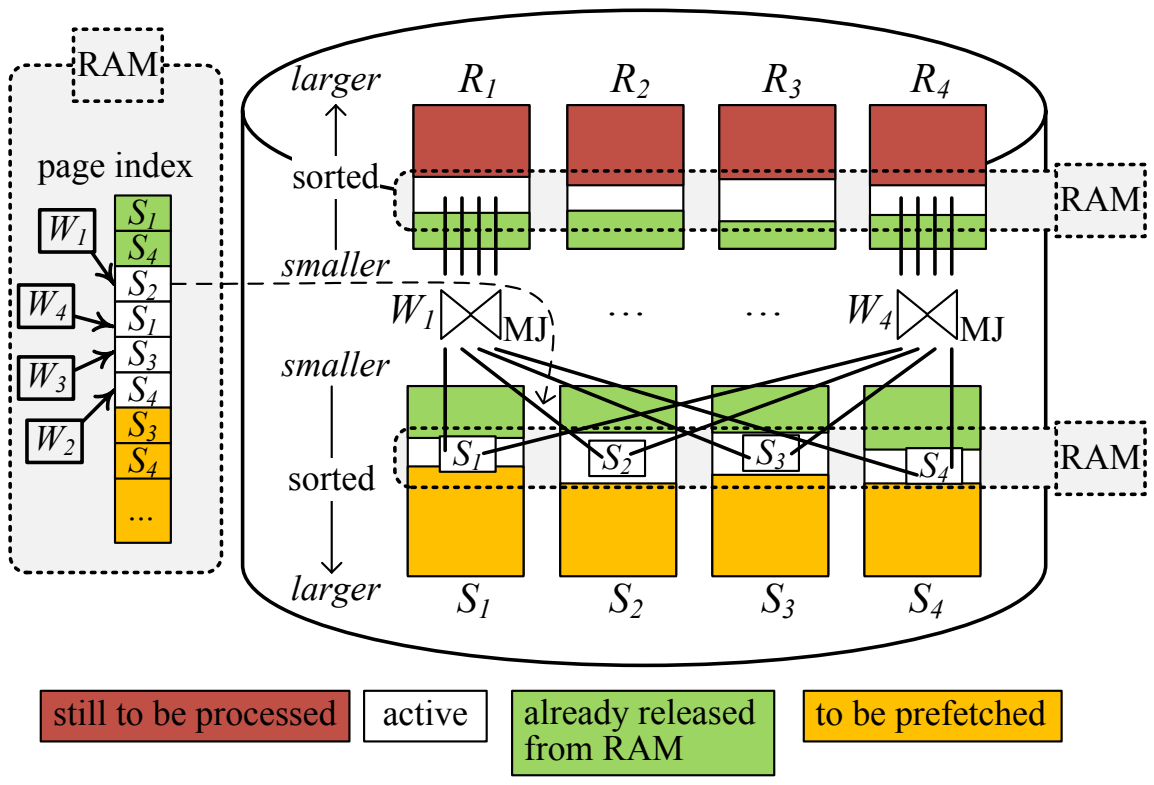

Figure 4: Disk-enabled MPSM join: the four workers $W_i$ progress synchronously through their $R_i$ run and all $S$ runs, thereby only active parts of the runs are in RAM

P-MPSM is a pure main memory version that range partitions the input data thereby providing scalability with respect to processing cores. D-MPSM is a RAM-constrained version that spools runs to disk. Both scenarios are common in main memory DBMSs and require attention when database operators are designed. We carefully consider both variants detailed enough to allow for an implementation and considerations about performance.

## 3.1 Memory-Constrained Disk MPSM

The presented MPSM can effectively be adapted to scenarios in which the intermediate result data is too large to be kept in main memory. Even in main memory database systems like HyPer that retain the entire transactional database in RAM, the query processor spools intermediate results to disk to preserve the precious RAM capacity for the transactional working set. Therefore, it is important to support both pure main memory algorithms and a disk-based processing mode with a very small RAM footprint.

The disk-enabled MPSM (D-MPSM) processes the left and right input runs by synchronously moving through the key domain which is sorted. The resulting data locality allows to spill already processed data to disk and to prefetch data that is to be processed soon. Figure 4 illustrates the approach: both $R$ and $S$ runs are stored on disk, only the currently processed pages (white) need to be main memory resident. Already processed data is not touched again and thus can be released from RAM (green) and soon to be processed data is prefetched from disk asynchronously (yellow).

For this purpose, we maintain a page index which is ordered page-wise by key value. The index is built during run generation and contains pairs $\langle v_{ij}, S_i \rangle$ where $v_{ij}$ is the first (minimal) join key value on the $j^{\text{th}}$ page of run $S_i$. Figure 4 depicts a simplified page index (only run identifiers) on the left. It actually contains the following information:

sorted by $v_{ij}$ →

| | | | | | | | | |
|---|---|---|---|---|---|---|---|---|
| $v_{11}$ | $v_{41}$ | $v_{21}$ | $v_{12}$ | $v_{31}$ | $v_{42}$ | $v_{32}$ | $v_{43}$ | ... |
| $S_1$ | $S_4$ | $S_2$ | $S_1$ | $S_3$ | $S_4$ | $S_3$ | $S_4$ | ... |

where $v_{11} \leq v_{41} \leq \ldots \leq v_{43}$. Both the prefetcher and the workers process the $S$ input data in the order specified by the index, thereby synchronously moving through the key domain and allowing to keep only a small part of the data in memory during join processing. All page index entries already processed by the "slowest" worker, e.g., $W_1$ in the illustration, point to run pages that may be released from RAM (green). The prefetcher is supposed to pre-load run pages according to the index before they are accessed by any worker (yellow). Implicitly, the workers' private input runs $R_i$ are read from disk (red), processed, and released from RAM in ascending order of join keys. Please note that the common page index structure does not require any synchronization as it is accessed read-only.

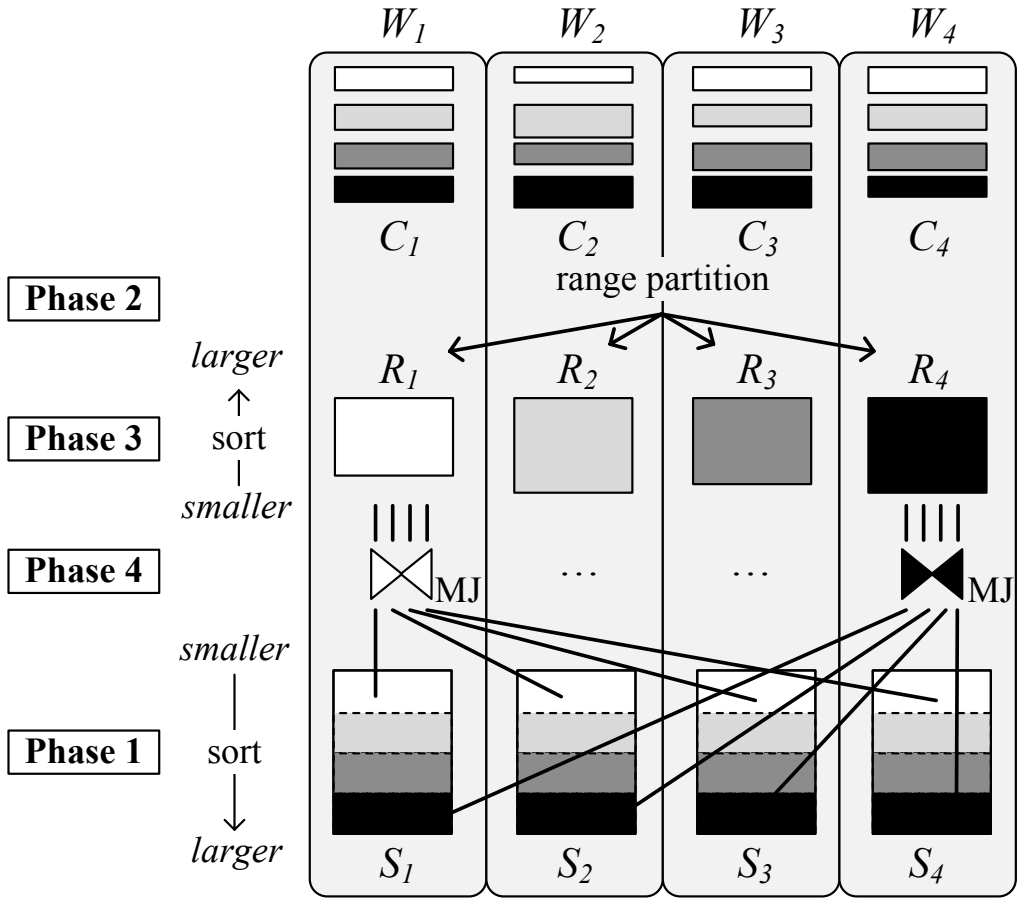

Figure 5: P-MPSM join with four workers $W_i$

Obviously, the performance of D-MPSM is determined by the time to write (run generation) and read (join phase) both inputs. Therefore, in order to exploit the power of multiple cores a sufficiently large I/O bandwidth (i.e., a very large number of disks) is required.

## 3.2 Range-partitioned MPSM

The range partitioned MPSM (P-MPSM) extends the B-MPSM algorithm by a prologue phase to range partition and assign the private input data to the workers in a way that allows saving much of the work during the join phase. The different phases of the algorithm are sketched in Figure 5 for a scenario with four workers, choosing $R$ as private input and $S$ as public input. In phase 1, the public input is chunked and sorted locally, resulting in runs $S_1$ to $S_4$. Subsequently, in phase 2, the private input is chunked into $C_1$ to $C_4$ and those chunks are range partitioned. We always employ a histogram-based technique to ensure that the range partitions are balanced (cf. Section 4) even for skewed data distributions. Thereby, the private input data is partitioned into disjoint key ranges as indicated by the different shades in Figure 5 ranging from white over light and dark gray to black. In phase 3, each worker then sorts its private input chunk and in phase 4, merge joins its own private run $R_i$ with all public input runs $S_j$.

By refining the MPSM to use range partitioning, each thread conducts only the join between $^1/_T{}^{\text{th}}$ of the join key domain of $R$ and $S$. This reduces the complexity per worker $W_i$ to

$$\begin{array}{rrl} & ^{|S|}/_T \cdot \mathit{log}(^{|S|}/_T) & \text{sort chunk } S_i \text{ of size } ^{|S|}/_T \\ + & ^{|R|}/_T & \text{range-partition chunk } R_i \\ + & ^{|R|}/_T \cdot \mathit{log}(^{|R|}/_T) & \text{sort chunk } R_i \text{ of size } ^{|R|}/_T \\ + & T \cdot {}^{|R|}/_T & \text{process run } R_i \text{ for all } S \text{ runs} \\ + & T \cdot {}^{|S|}/_{T^2} & \text{process } ^1/_T{}^{\text{th}} \text{ of each } S \text{ run} \end{array}$$

$$= {}^{|S|}/_T \cdot \mathit{log}(^{|S|}/_T) + {}^{|R|}/_T + {}^{|R|}/_T \cdot \mathit{log}(^{|R|}/_T) + |R| + {}^{|S|}/_T$$

Compared to the complexity approximation of B-MPSM, range partitioning pays off if the cost of range-partitioning $R$ is smaller than the savings in join processing, i.e., if

$$^{|R|}/_T \leq |S| - {}^{|S|}/_T$$

For a parallelism greater than or equal two and $|R| \leq |S|$ it pays off. The performance of P-MPSM thus scales almost linearly with the number of parallel threads $T$ which is decisive for the effective multi-core scalability of P-MPSM, as our experimental evaluation will also prove.

In general, the two input relations to a join operation are not equally sized but usually consist of a larger (fact) table and smaller (dimension) tables. Assigning the private input role $R$ to the smaller of the input relations and thus the public input role $S$ to the larger yields the best performance. Thereby, only a small fraction (depending on the number of worker threads $T$) of the remote public input needs to be processed while the smaller private input is scanned several times with almost no performance penalties. We will present evaluation results quantifying the performance impact of reversed public/private input roles in Section 5.4.

### 3.2.1 Partitioning the Private Input (Phase 2)

We design the re-distribution of the private input chunks $C_i$ to be very efficient, i.e., *branch-free*, *comparison-free*, and *synchronization-free*. (1) *branch-freeness* and *comparison-freeness* are achieved by using radix-clustering [19] on the highest $B$ bits of the join key where $\mathit{log}(T) \leq B$. For $\mathit{log}(T) = B$, radix-clustering results in exactly $T$ clusters. By increasing $B$, we can account for skew in both $R$ and $S$ as we will discuss in Section 4. (2) We then range partition the private input chunks, thereby guaranteeing *synchronization-freeness* by letting each worker write sequentially to precomputed sub-partitions within all runs. For this purpose, each thread builds a histogram on its chunk of the global relation $R$. The local histograms are combined to obtain a set of prefix sums where each prefix sum represents the start positions of each worker's partitions within the target runs. Each worker then scatters its input chunk to the partitions using the prefix sums and updating them accordingly. This approach was adapted from the radix join of [14].

We demonstrate the partitioning of $R$ in Figure 6 for two workers, $B = 1$ and a join key range of $[0, 32)$. Each worker thread $W_i$ scans its own chunk $C_i$ and probes for each tuple into a histogram array depending on its highest bit (which we show underlined), i.e., join key values $< 16$ are assigned to the first position and join key values $\geq 16$ are assigned to the second. According to $h_1$, chunk $C_1$ contains four entries for the first and three for the second partition. The two partitions are shown as white and black entries. From the combined histograms prefix sums are computed that point to the subpartition into which the workers scatter their chunk's tuples. For example, the prefix sum $ps_1$ denotes that $W_1$ scatters its entries for the first and second partition starting at position 0. According to $ps_2$, $W_2$ scatters tuples belonging to the first partition beginning at position 4 (as $W_1$ writes to positions 0 to 3), and those belonging to the second partition beginning at position 3. In general, the $ps$-entries are computed as

$$ps_i[j] = \begin{cases} 0, & \text{if } i = 1 \\ \sum_{k=1}^{i-1} h_k[j], & \text{else} \end{cases}$$

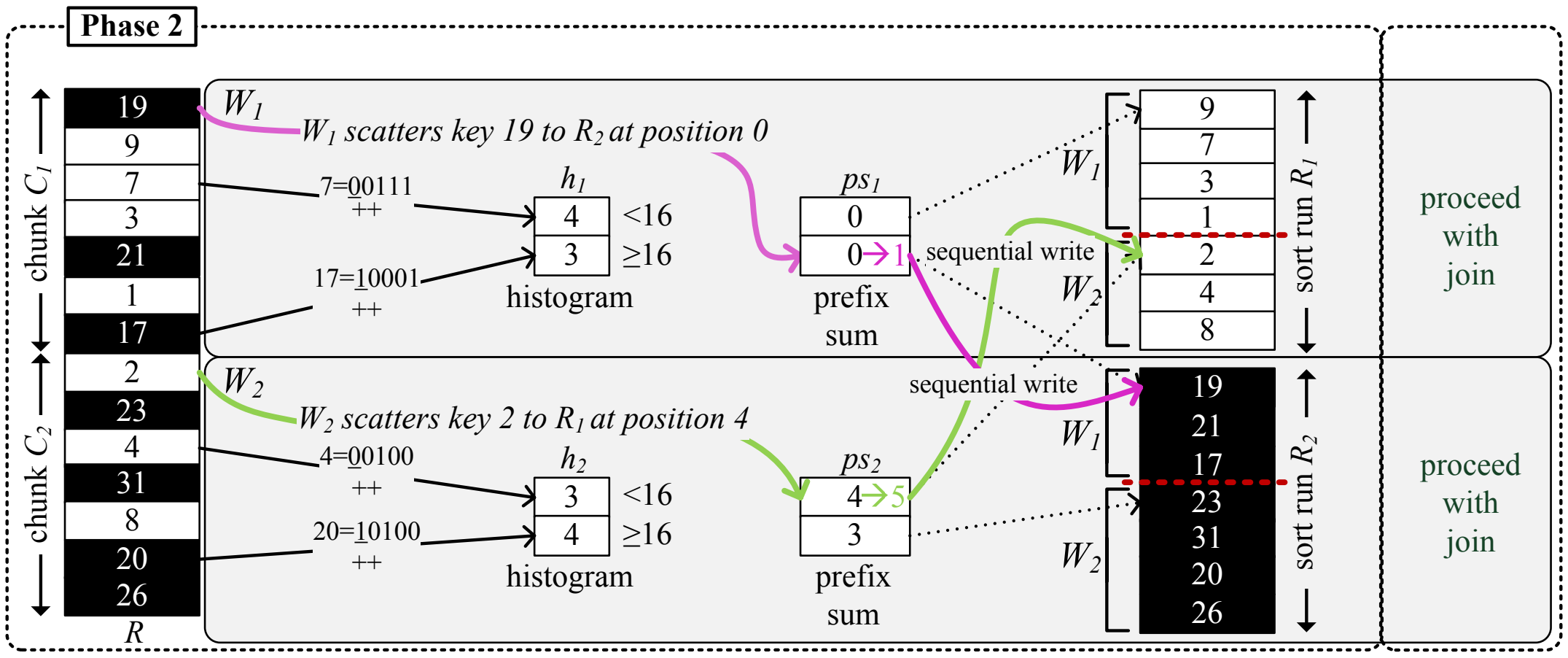

Figure 6: Phase 2 of P-MPSM: 5 bit join keys in the range [0,32), 1 bit histogram for 2 partitions

Actually, the $ps_i$ contain pointers to the positions, not index values, as shown by the dotted arrows in Figure 6, i.e., $ps_i[j] = \&R_j[(\sum_{k=1}^{i-1} h_k[j])]$. The prefix sums $ps_i$ per worker $W_i$, which are computed from the combined local histograms, are essential for the synchronization-free parallel scattering of the tuples into their range partition. Every worker has a dedicated index range in each array $R_i$ into which it can write sequentially. This is orders of magnitude more efficient than synchronized writing into the array – as shown in Figure 1 (2) and makes MPSM immune against cache coherency overhead.

Note that depending on the actual join key value distribution, in particular the minimum and maximum join key values, it might be necessary to preprocess the join keys before applying radix-clustering. This can usually be done efficiently using bitwise shift operations.

Although we use radix-clustering for partitioning the private input, the approach is not restricted to integer join keys. However, if long strings are used as join keys, MPSM should work on the hash codes of those strings, thereby giving up the meaningful sorting of the output. Furthermore, main memory DBMSs usually employ dictionary encoding so that joins on strings are usually internally executed as joins on integers anyway.

### 3.2.2 *Join Phase (Phase 4)*

Due to partitioning, the private input data chunks contain only a fraction of the key value domain and thus probably join only with a fraction of each public input data chunk. As indicated in Figure 5, the public input runs are therefore **implicitly partitioned** – by the sorting order. Sequentially searching for the starting point of merge join within each public data chunk would incur numerous expensive comparisons. Thus, we determine the first public input tuple of run $S_j$ to be joined with the private input run $R_i$ using *interpolation search* as sketched in Figure 7.

Depending on which of the first values of each run – $s_{j1}$ and $r_{i1}$ – is larger (in general this will be $r_{i1}$ because the key range of $R$ runs is limited while the key range of $S$ runs is not), we search for it within the other run by iteratively narrowing the search space. The most probable position in each iteration is computed by applying the rule of proportion using the minimum and maximum index positions and the minimum and maximum key values of the current search space, as well as the difference of the searched key value and the search space minimum key value. The computed index per iteration is always relative to the search space starting index. In the illustration in Figure 7, only two steps are required to find the starting point for merge join:

1. the search space is $[s_{j1}, s_{jn}]$, i.e. from indexes 1 to $n$, thus we compute $1 + (n-1) \cdot {}^{(r_{i1} - s_{j1})}/_{(s_{jn} - s_{j1})} = k$
2. the search space is narrowed to $[s_{j1}, s_{jk}]$, i.e. from indexes 1 to $k$, so we compute $1 + (k-1) \cdot {}^{(r_{i1} - s_{j1})}/_{(s_{jk} - s_{j1})}$ and find the start index of the light gray partition.

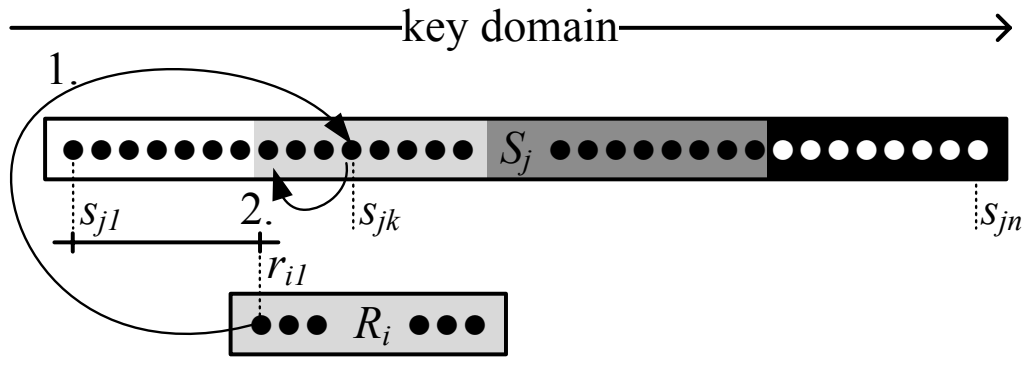

Figure 7: Interpolation search for $r_{i1}$ in $S_j$

## 4. SKEW RESILIENCE OF P-MPSM

The basic B-MPSM as well as the disk variant D-MPSM are completely skew immune as they do not range partition.

So far, we discussed P-MPSM using statically determined partition bounds. In case of uniform data distribution the presented algorithm assigns balanced workloads (i.e., equally sized chunks) to the workers. It is important to note that the location of the data – e.g., if by time of creation clustering small values appear mostly before large values – within the relations $R$ and $S$ has no negative effect. The *location skew* among the $R$ and $S$ runs is implicitly handled by range partitioning the $R$ data and thereby limiting the $S$ data each worker has to process. Of course, location skew may cause slight NUMA effects that cannot be controlled lastly. As our evaluation in Section 5.5 shows, these effects usually have a positive impact on performance as the join partners of a partition $R_i$ are better clustered in $S$.

We now present a more elaborate version of P-MPSM that can handle *distribution skew* while incurring only very little overhead to the overall performance. Skew resilience is achieved by not determining the partition bounds statically but computing them based on dynamically obtained information about the key value distributions in $R$ and $S$. We exploit the sort order of the public input $S$ to compute arbitrarily precise histograms representing the key value distribution of $S$ en passant, i.e., in almost no time. Further, we increase the number $B$ of bits used for the histogram computation for radix-clustering of the private input $R$ and thereby also obtain very precise histograms representing the private input join key value distribution. We then determine global load-balancing partition bounds based on the computed distributions. We show that the presented approach adds only very little overhead to the overall join processing.

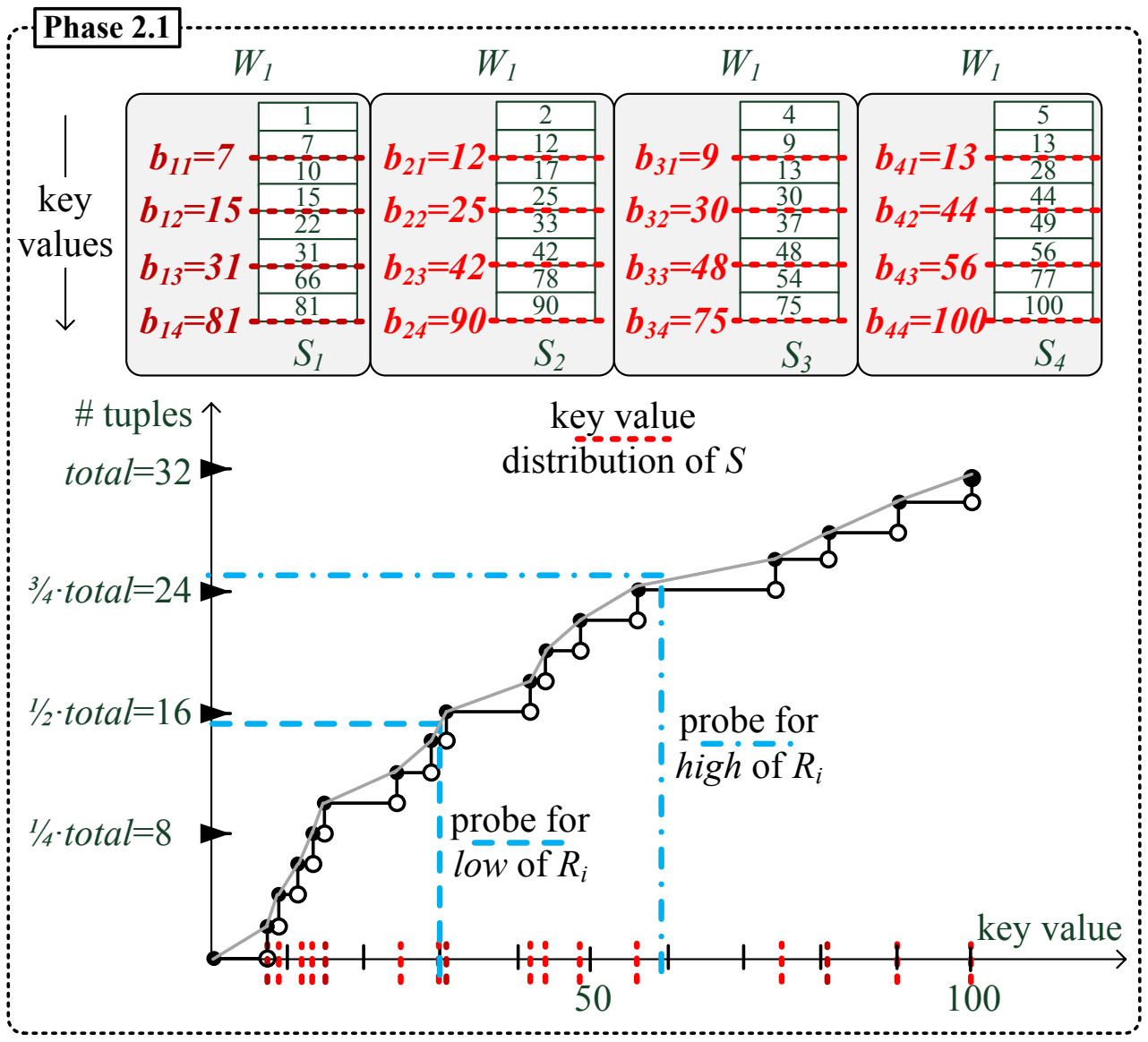

Figure 8: P-MPSM CDF computation: example with skewed input (mostly small key values)

For better illustration, we split the range partition phase 2 into the following sub-phases: The histogram on $S$ is determined in phase 2.1 using a cumulative distribution function (CDF). The histogram on $R$ is determined in phase 2.2 using probing as described above but increasing the number of leading bits $B$ used for fine-grained histogram boundaries. In phase 2.3 we combine the information about the key value distributions in $R$ and $S$ to find global partition bounds, called splitters, balancing the costs for sorting $R$ chunks and joining. This way, we ensure that each worker thread is assigned a balanced workload to make sure that they all finish at the same time which is very important for subsequent query operators.

## 4.1 Global $S$ Data Distribution (Phase 2.1)

We gain insight in the global $S$ data distribution in two steps: First, each worker thread $W_i$ computes an equi-height histogram for its local input run $S_i$. Building the equi-height histograms comes at almost no costs as the data is already sorted. Then, the local histograms are merged to provide a global distribution view. The procedure is exemplified in Figure 8 for four runs $S_1$ to $S_4$ with skewed data, i.e., small join key values occur much more often than large join key values. The local equi-height histogram bounds $b_{ij}$ for each worker $W_i$ computed during the first phase are marked as red dotted lines within the input runs. In the example, each worker collects four local bounds, i.e., the local histograms are of size four. In the second phase, the local partition bounds of all workers are collected as input to a global cumulative distribution function (CDF).

Using the local equi-height histograms we can only estimate the gradient of the step function by approximating each step to be equally high. Of course, the real global distribution deviates (slightly) from this as the different workers' equi-height partitions have overlapping key ranges. In the example in Figure 8, each worker thread determines $T = 4$ local bounds, in general we propose to compute $f \cdot T$ local bounds for better precision. By increasing $f$ and thus the number of local bounds determined by each worker, more fine grained information about the global data distribution can be collected at negligible costs.

Note that the CDF allows for configuration changes concerning the number of workers. Appropriate partition limits are then found using interpolation as denoted in Figure 8 by the diagonal connections between steps. This also allows to combine the global $S$ data distribution represented by the CDF with the $R$ data distribution in order to handle uncorrelated or even negatively correlated skew in $R$ and $S$ as we will show below.

## 4.2 Global $R$ Distribution Histogram (Phase 2.2)

In phase 2.2, each worker scans its private input chunk $C_i$ and computes a local histogram on it using radix-histogramming. Thereby, the number of leading bits $B$ determines the precision of the histogram, i.e., using $B$ bits we obtain a histogram of size $2^B$. Building a more fine-grained histogram does only incur little overhead but allows for a much more precise computation of global $R$ bounds. By merging some of the clusters to form $T$ partitions with a balanced workload ($cost(sort(R_i))+cost(R_i \bowtie S)$) we obtain the global partition bounds. On the left hand side of Figure 9 we see that higher precision of radix-histogramming comes at no additional cost. On the right hand side we see the inferior performance of comparison-based partitioning.

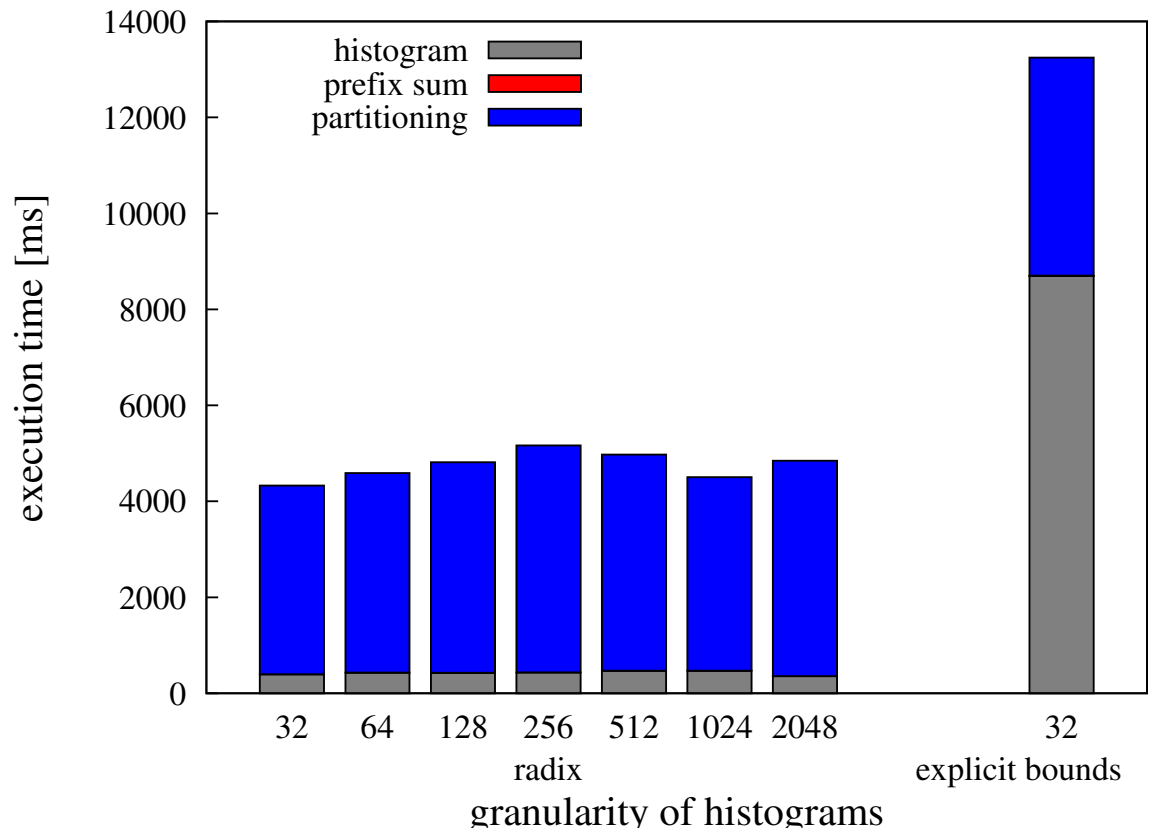

Figure 9: Fine-grained histograms at little overhead

In Figure 10 the proceeding is exemplified for a scenario with two workers clustering two initial chunks and redistributing them to two target partitions. They first build a local histogram of size 4 ($B = 2$) each, dividing the skewed input data with key domain $[0, 32)$ into four partitions: $< 8$, $[8, 16)$, $[16, 24)$, $\geq 24$. The histograms reveal that the chunks contain many more small key values than large key values, in particular, there are a total of seven values in the first partition, three values in the second, three values in the third, and one value in the fourth partition.

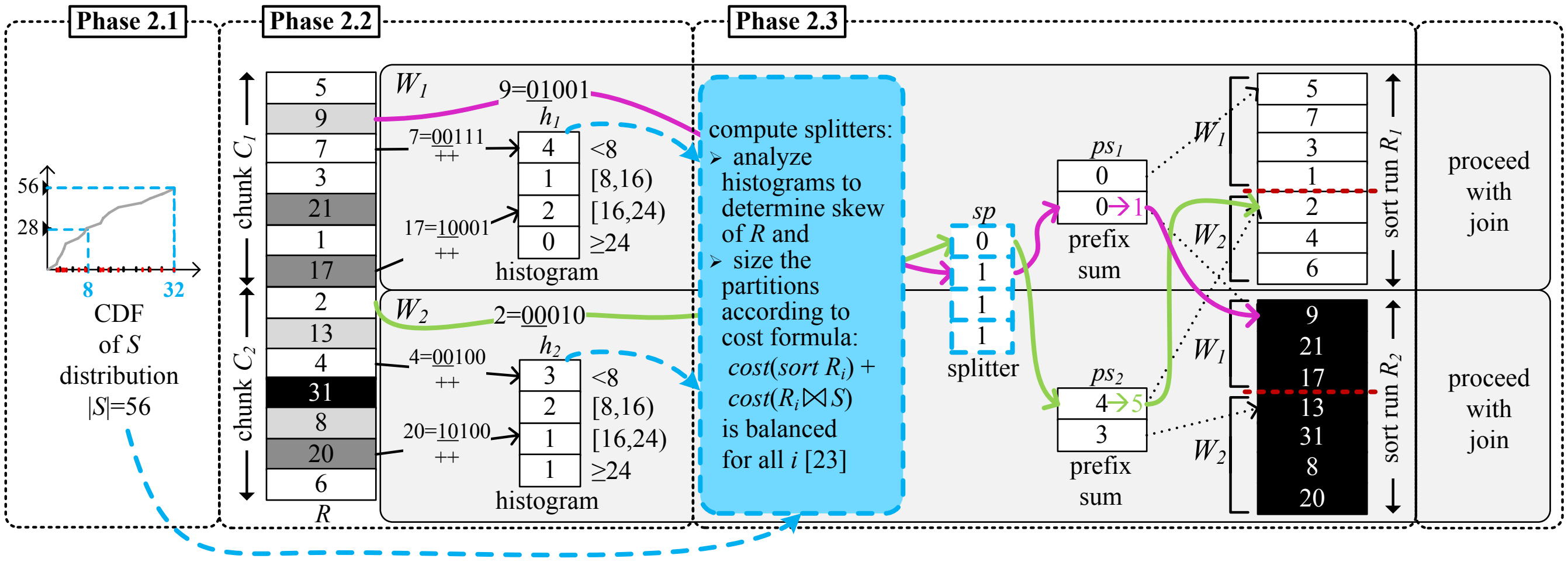

Figure 10: Load balanced partitioning of the private input $R$: join keys in the range $[0, 32)$ are skewed (mostly small)

## 4.3 Partitioning the Private Input $R$ (Phase 2.3)

We use the global CDF for $S$ determined in phase 2.1 and the global $R$ distribution histogram from phase 2.2 to heuristically determine global partition bounds using a complexity approximation that takes into account both the sort costs of the $R$ chunks and the join costs per worker $W_i$:

$$\begin{array}{lrl} \multicolumn{3}{l}{\textit{split-relevant-cost}_i =} \\ & |R_i| \cdot \textit{log}(|R_i|) & \text{sort chunk } R_i \\ + & T \cdot |R_i| & \text{process run } R_i \\ + & \text{CDF}(R_i.\textit{high}) - \text{CDF}(R_i.\textit{low}) & \text{process relev. } S \text{ data} \end{array}$$

where $R_i.low$ and $R_i.high$ denote the radix boundaries for which we probe in the CDF. Note that because of the sorting $S$ can be partitioned at any position. The boundaries are determined at the radix granularity of $R$'s histograms.

As shown in Figure 8 and on the left of Figure 10 using blue dashed lines, the tentative $R$ histogram bounds are used to probe into the CDF to determine the anticipated $S$ costs for the currently considered $R$ partition $[low, high)$. If the key values in $R$ and $S$ are uniformly distributed or skewed in a correlated way, the global $R$ partition bounds will be similar to the global $S$ partition bounds and thus all $R_i$ will be approximately equally sized. If the key value distribution is uncorrelated, they may be very different so that we need to weight their effect on the overall performance to find the final global partition bounds.

We opt to partition $R$ and $S$ such that each worker is assigned the same amount of work, i.e., we determine the partition bounds such that they minimize the biggest cost $\textit{split-relevant-cost}_i$ over all $1 \leq i \leq T$. We refer to Ross and Cieslewicz [23] who present elaborate techniques for finding optimal partition bounds for two table partitioning problems.

In the example in Figure 10, to simplify matters we assume the key value distribution of $S$ to be correlated to that of $R$. Therefore, when probing into the CDF using 8 and 32 as $[low,high)$ values for $R_2$, those bounds divide $S$ in equally sized partitions. Thus, according to the histograms $h_1$ and $h_2$ and the CDF of $S$, the first cluster $< 8$ becomes the first partition and the other three clusters $\geq 8$ form the second partition.

We then partition the private input chunks using the global partition bounds. Thereby, we avoid synchronization by letting each worker write sequentially to precomputed partitions. For this purpose, the local histograms are combined to a set of prefix sums where each prefix sum represents the workers' partitions within the target runs. In phase 2.3, each worker scatters its input chunk to the partitions using the prefix sums via the indirection of the splitter vector $sp$, i.e., worker $W_i$ scatters its next tuple $t$ as follows:

$$\texttt{memcpy}(ps_i[sp[t.key \gg (64-B)]]\texttt{++}, t, t.size)$$

$ps_i$ contains pointers, not indexes because each worker scatters to different arrays. According to the global $R$ partition bounds $b_1 = 8$ and $b_2 = 32$, there are four values of chunk $C_1$ falling into the first partition and three (1+2+0) falling into the second. From chunk $C_2$, three values belong to the first and four (2+1+1) to the second partition. The local histograms (which are computed per chunk) are combined to global prefix sums. The values in $ps_1$, for instance, denote that worker $W_1$ should scatter its data falling into the first partition to run $R_1$ beginning at position 0, whereas worker $W_2$ should write its data for the first partition to run $R_1$ beginning at position 4. Thereby, $ps_i$ is incremented for each tuple scattered. Please note that – depending on the key distribution in $R$ – the resulting runs might not be of equal size. It is more important that the cost is balanced rather than the size (cf. Section 5.6). Unlike radix join, MPSM can partition the private input $R$ completely independently of $S$. The public input $S$ is partitioned implicitly via the sorting and thus does not incur any partitioning overhead.

# 5. EXPERIMENTAL EVALUATION

We implemented the MPSM join variants in C++, and the join query plans are compiled as employed in our HyPer query processor [21]. All experiments are such that the data is completely in main memory. For the disk-enabled Vectorwise this is achieved by executing the query several times and reporting only the execution times of runs after the data was fully resident in RAM. In order to cover the most important scenarios, we report benchmark results using datasets representing join input relations of different sizes, different multiplicities, and different data distributions. We consider the common case that two relations $R$ and $S$ are scanned, a selection is applied, and then the results are joined. So, no referential integrity (foreign keys) or indexes could be exploited.

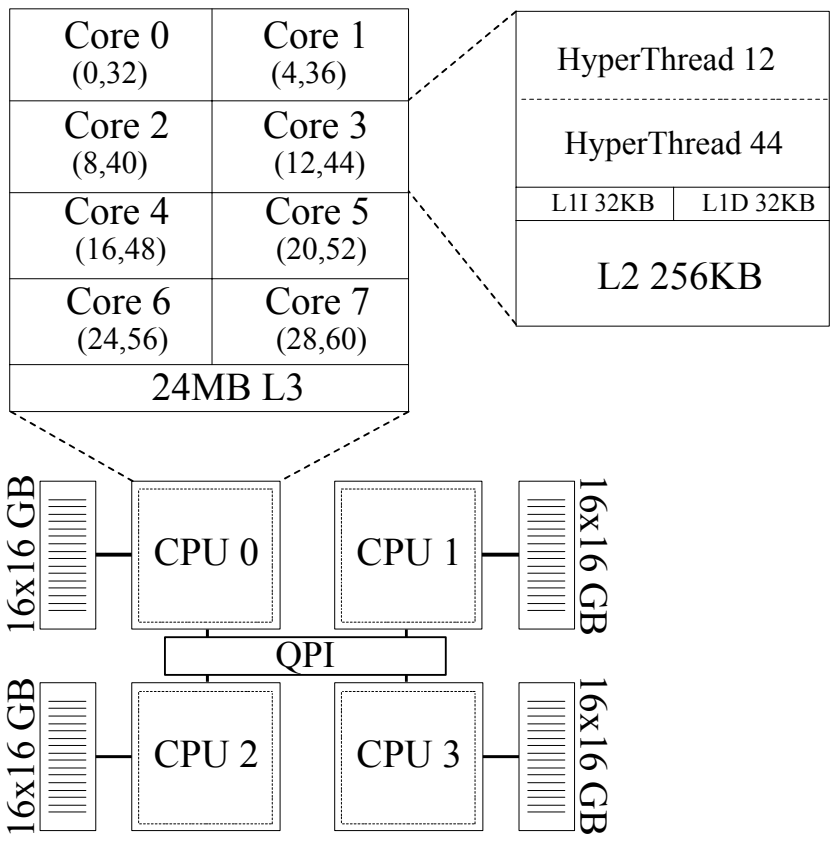

Figure 11: Intel 32 core, 1 TB server (HyPer1)

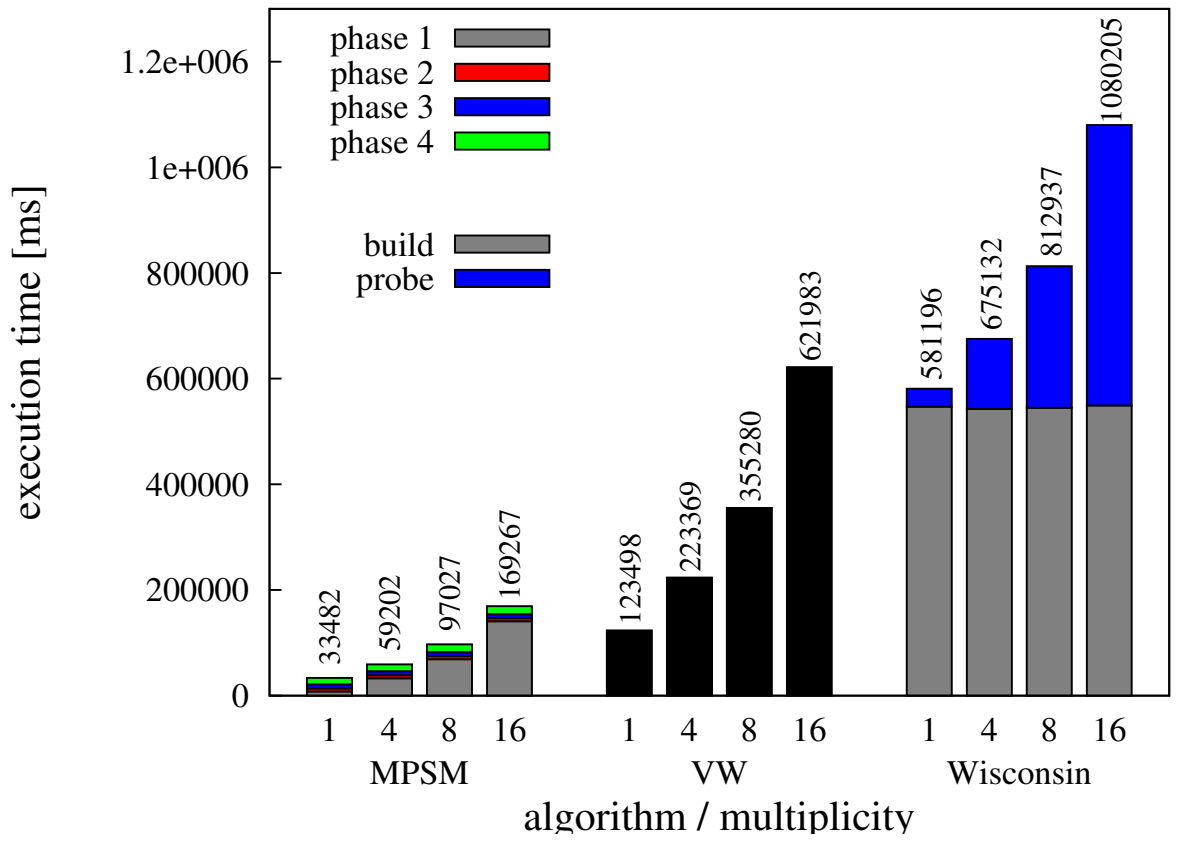

Figure 12: MPSM, Vectorwise (VW), and Wisconsin hash join on uniform data

Due to space limitations we will concentrate our experimental evaluation on the in-memory range-partitioned variant P-MPSM and leave the analysis of the disk variant for future work.

## 5.1 Platform and Benchmark Scenarios

We conduct the experiments on a Linux server (kernel 3.0.0) with one TB main memory and four Intel(R) Xeon(R) X7560 CPUs clocked at 2.27GHz with 8 physical cores (16 hardware contexts) each, resulting in a total of 32 cores (and due to hyperthreading 64 hardware contexts) as in Figure 11. This machine has a list price of approximately €40000 which makes it a good candidate for the real time business intelligence scenario on transactional data for which our HyPer main memory DBMS is intended (therefore the machine is called HyPer1 in our lab).

As "contenders" we chose the most recent research system, published by the Wisconsin database group [1] and the "cutting-edge" Vectorwise query engine which holds the world record in single-server TPC-H power test. According to our tests it currently has the best-performing parallel join processing engine which is based on the pioneering MonetDB work on cache-friendly radix joins [19]. This is also testified by Vectorwise's record TPC-H powertest performance on "small" main memory fitting databases up to one TB on a single machine. (Actually, the TPC-H record numbers were obtained on a similar machine as our HyPer1.) For the SIGMOD2011 Wisconsin hash join benchmarks we use the original code [1]. The Vectorwise benchmarks were conducted on Vectorwise Enterprise Edition 2.0.

We chose the datasets to be representative for a few realistic data warehouse scenarios. Each dataset consists of two relations $R$ and $S$. The cardinality of $R$ is $1600M$, the cardinality of $S$ is scaled to be $1 \cdot |R|$, $4 \cdot |R|$, $8 \cdot |R|$, and $16 \cdot |R|$. These database sizes are one order of magnitude larger than in prior related studies [17, 1, 2] to account for recent hardware improvements in RAM capacity and real-world requirements in operational business intelligence. For example, Amazon has a yearly revenue of $40 billion, for which an estimated item price of $10 results in 4 billion orderline tuples – a size which is covered by our experiments. It is interesting to note that the transactional sales data of this largest merchandiser, if properly normalized and possibly compressed, fits into the RAM of our one TB machine which makes operational BI on main memory resident data a reality – if the parallelization power of these machines can be effectively exploited.

Each tuple consists of a 64-bit key within the domain $[0, 2^{32})$ and a 64-bit payload:

$$\{[joinkey\text{: 64-bit}, payload\text{: 64-bit}]\}$$

We execute an equi-join on the tables:

```
SELECT max(R.payload + S.payload)
FROM R, S
WHERE R.joinkey = S.joinkey
```

This query is designed to ensure that the payload data is fed through the join while only one output tuple is generated in order to concentrate on join processing cost only. Further, we made sure that early aggregation was not used by any system. We chose the data format both for scaling reasons (payload may represent a record ID or a data pointer) as well as for ease of comparison reasons to the experiments presented in [1]. Our datasets of cardinality $1600M\times$ (1+multiplicity) have sizes ranging from 50 GB to 400 GB which is representative for large main memory operational BI workloads. The multiplicities between the relations $R$ and $S$ further cover a wide range, including not only the common cases (4, as specified for instance in TPC-H and 8 to approximate the TPC-C specification) but also extreme cases (1 and 16). We further experimented with skewed datasets.

## 5.2 Comparison of MPSM, Vectorwise, and Wisconsin Join on Uniform Data

We compare MPSM, Vectorwise, and Wisconsin join on uniform data for different multiplicities (in the extreme case $S$ is 16 times as large as $R$). The results are shown in Figure 12. MPSM outperforms Vectorwise by a factor of four. Wisconsin is not adapted to efficiently work for NUMA architectures as it builds and probes a global hash table across NUMA partitions, which results in poor performance for such large data volumes and numbers of cores. Therefore, we don't consider it in further experiments.

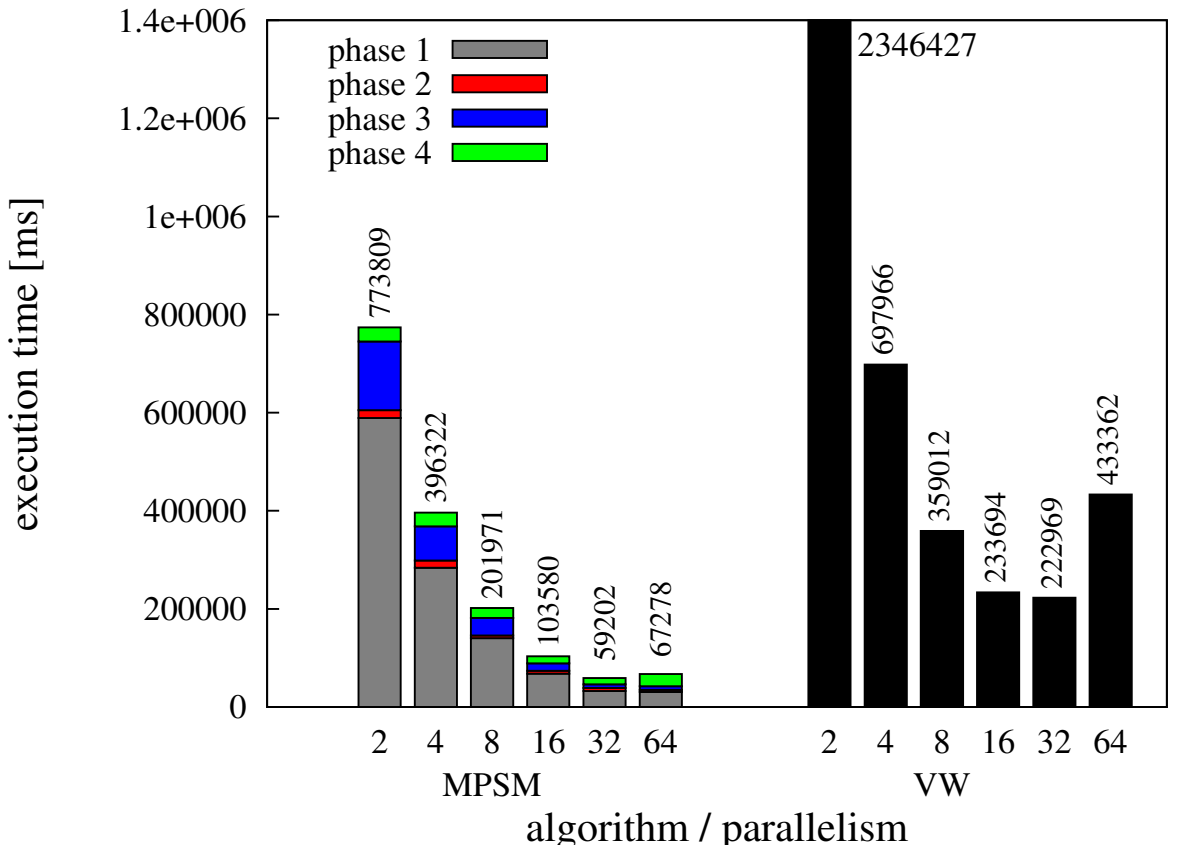

Figure 13: Scalability of MPSM and Vectorwise

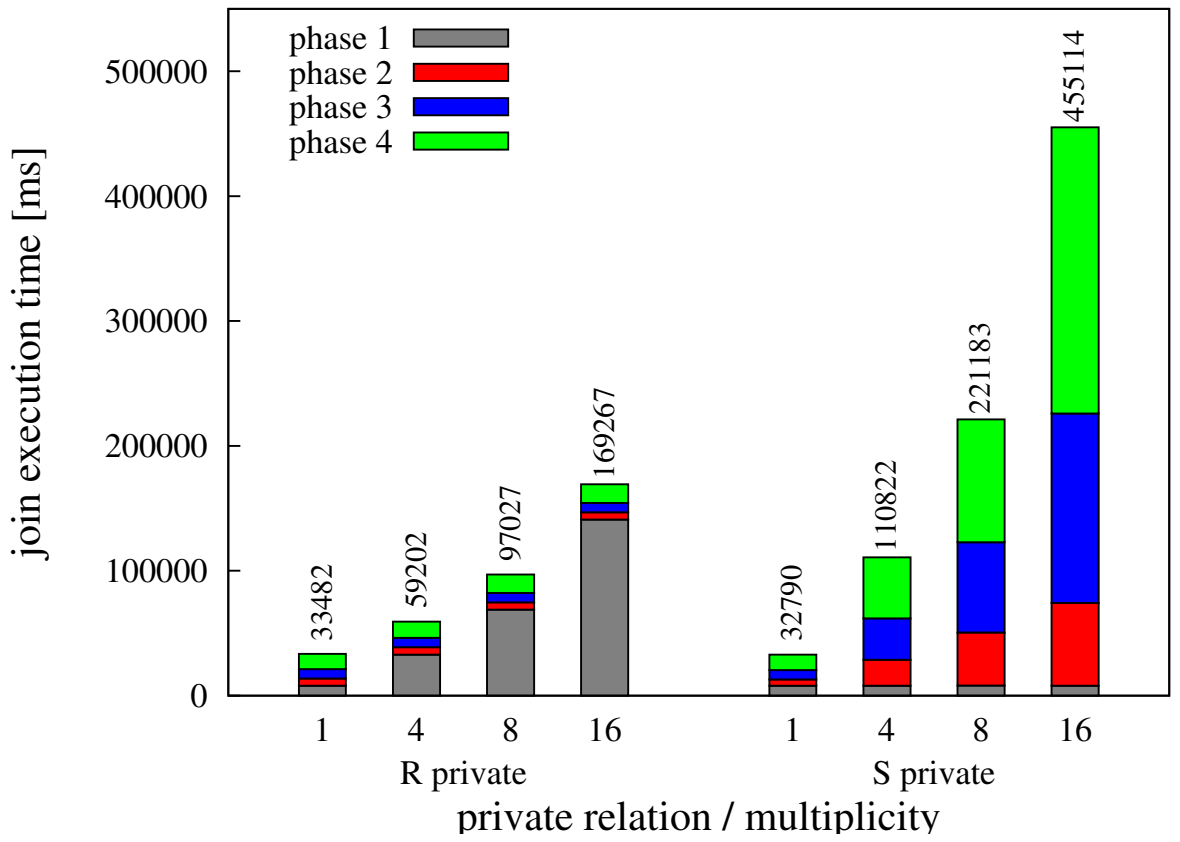

Figure 14: Effect of role reversal on join execution time

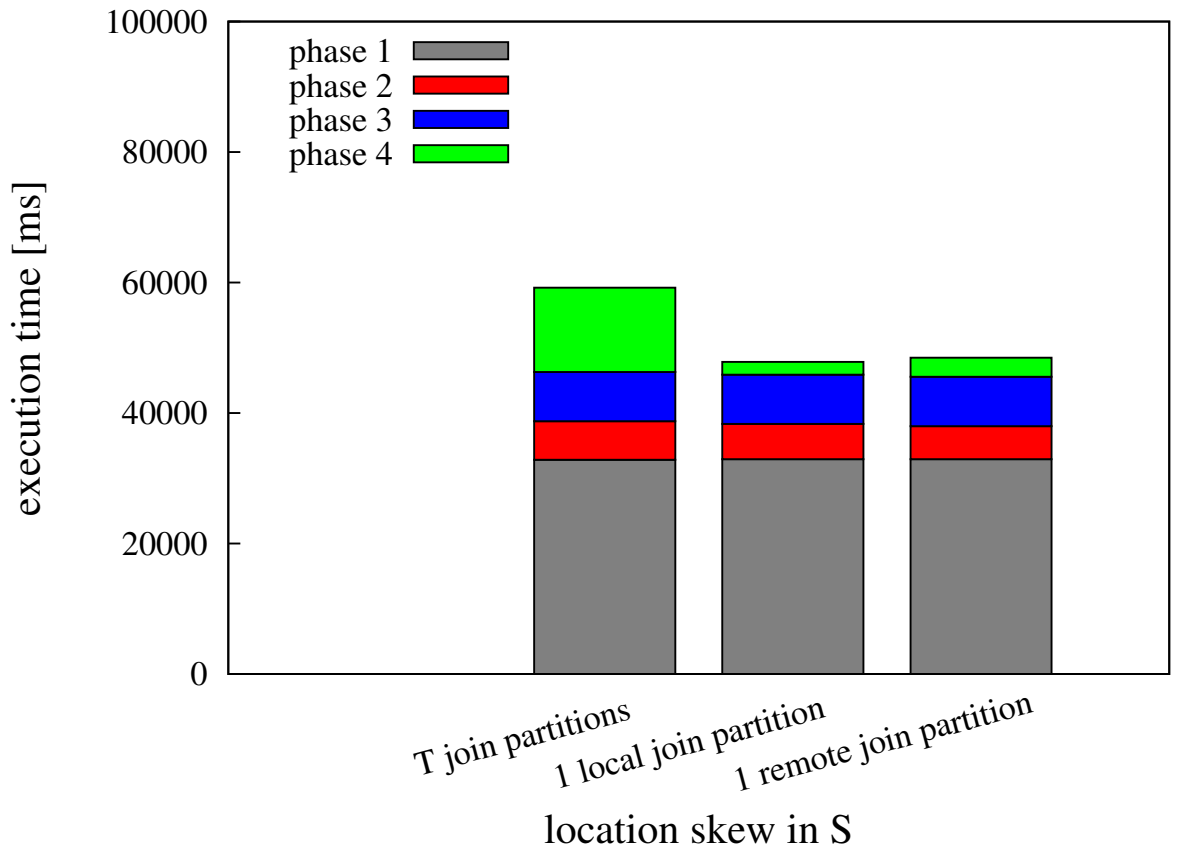

Figure 15: Location skew (32 workers, multiplicity 4)

## 5.3 Scalability in Number of Cores

We compare the scalability with respect to the number of cores for MPSM and Vectorwise and report the results in Figure 13. MPSM scales almost linearly in the number of parallel executing worker threads. As depicted in Figure 11, our server has 32 physical cores and a total of 64 hardware contexts. When exceeding the number 32 of physical cores and using hyperthreading (parallelism level 64), the performance of MPSM remains stable but does not improve as all cores are already fully utilized at parallelism 32. From these results, we are confident, that MPSM will scale well on future hardware with even hundreds of cores.

## 5.4 Role Reversal

We mentioned in Section 3.2 that for P-MPSM it is advisable to consider role reversal for performance improvements. In Figure 14 we compare the execution time for two relations $R$ and $S$ where we vary the size of $S$ to be *multiplicity* times that of $R$. Thereby, phase 1 (sorting the public input) and phase 3 (sorting the private input) are interchanged and have the same execution time when summed up. However the effect of role reversal is clearly visible for the range partition phase 2 and the join phase 4. For the multiplicity 1, role reversal obviously has no effect on the join execution time (as both inputs have the same size). However, the larger $S$ grows, the more considerable is the effect that directly follows from $|R| < |S|$ and the complexity estimate in Section 3.2 (ignoring the equal sort costs) as

$$|R|/T + |R| + |S|/T < |S|/T + |S| + |R|/T$$

## 5.5 Location Skew

We introduced location skew by arranging $S$ in small to large join key order – no total order, so sorting the clusters was still necessary. Location skew on $R$ has no effect at all as $R$ is redistributed anyway. Extreme location skew of $S$ means that all join partners of $R_i$ are found in only one $S_j$. This results in each worker $W_i$ effectively producing join results only with one local $S_i$, respectively one remote $S_j$ where $i \neq j$. This is the extreme case as only/no local $S$ data contributes to the join result and only one remote memory area has to be accessed. Of course, all runs are still accessed using interpolation search, however, no relevant data is found in $(T-1)$ of the $S$ runs. This effectively reduces the complexity from

$$|S|/T \cdot log(|S|/T) + |R|/T + |R|/T \cdot log(|R|/T) + |R| + |S|/T$$

to

$$|S|/T \cdot log(|S|/T) + |R|/T + |R|/T \cdot log(|R|/T) + |R|/T + |S|/T$$

as the private $R_i$ is only scanned once to produce all join results. If there is less pronounced location skew in $S$, the algorithm performance lies between those two extremes shown in Figure 15. Note that in all other experiments location skew was not present/exploited.

## 5.6 Skewed Data with Negative Correlation

In this sort of experiments we analyze the quality of the splitter computation (cf. Figure 10) to balance the load evenly across all workers. For this purpose we generated a dataset with the worst possible skew for our join algorithm: negatively correlated skew in $R$ and $S$. (Positively correlated skew does not affect MPSM either due to the dynamic splitter computation.) Our data set again contained $1600M$ tuples in $R$ with an 80:20 distribution of the join keys: 80% of the join keys were generated at the 20% high end of the domain. The $S$ data of cardinality $4 \cdot 1600M$ was generated with opposite skew: 80% of the join keys at the low 20% end of the domain. Let us refer to Figure 16a to intuitively explain the necessity of balancing the load according to the data distribution of $R$ **and** $S$. On the left-hand side we show the effects of partitioning $R$ into equal-cardinality partitions thereby having wider ranges on the left and narrower ranges on the right. Because of the negative correlation the corresponding $S$ partitions are very unbalanced – so the combined cardinality of the two partitions $|R_i| + |S_{R_i}|$ is much higher at the low end than at the high end. $S_{R_i}$ denotes the relevant join range of $S$ for the range of $R_i$. Note that $S_{R_i}$ is composed of sub-partitions across all $S_1, \cdots, S_T$ but its size can effectively be estimated from the CDF. For 32 workers operating on this equi-height $R$ partitioning we obtain the response times shown in Figure 16b. We see that the "blue" sort costs are balanced but the "green" join processing takes much longer for the workers on the left that process the low join keys. The correct splitter-based partitioning balances the load across all servers – as shown in Figure 16c. The figure is idealized in balancing the cardinality of the two corresponding partitions; in reality the sort+join **costs** are balanced. This is achieved by considering the cardinality of each $R_i$ in combination with its corresponding $S_{R_i}$ partition which is obtained from the CDF. This balanced *R-and-S* partitioning is visualized in Figure 16a on the right hand side. For this experiment we computed the $R$ histograms at a granularity of 1024 (B=10) to give the splitter computation sufficient opportunity to find best possible splitters.

## 5.7 Experiments' Summary

We showed that MPSM scales to very large data volumes and scales almost linearly to increasing number of cores. This indicates MPSM's future potential as the upcoming servers will have even more RAM and more cores – probably a few hundred soon to come.

The superior performance characteristics of MPSM are corroborated by the observation that, in our experiments we neither

1) exploited any possibly existing sort order nor

2) exploited the quasi-sorted'ness of the result

In complete QEPs both aspects would favor the MPSM join even more in comparison to hash-based variants.

# 6. RELATED WORK

Parallel join processing originates from the early work on database machines, e.g., Gamma [9], where hash-based partitioning was used to distribute the join argument to multiple machines in a compute cluster. In this context some heuristics for skew handling were developed [8]. Teubner et al. [11, 24] present parallel joins for modern distributed databases. In multi-core parallel processing the distribution of the data is much more efficient as we can exploit the shared memory, albeit regarding the consequences of the NUMA architecture [22]. Our MPSM join is, to the best of our knowledge, the first work that consequently takes NUMA into consideration, which is decisive for large scale in-core databases. Most previous approaches to in-core parallel join processing were based on the radix join prioneered by the MonetDB group [19, 3]. This join method achieves cache locality by continuously partitioning into ever smaller chunks that ultimately fit into the cache. Ailamaki et al. [5] improve cache locality during the probing phase of the hash join using software controlled prefetching. Our sort-based MPSM algorithm has high cache locality and hardware prefetcher affinity by its very own merge join behavior that sequentially scans a pair of runs.

An Intel/Oracle team [17] adapted hash join to multi-core CPUs. They also investigated sort-merge join and hypothesized that due to architectural trends of wider SIMD, more

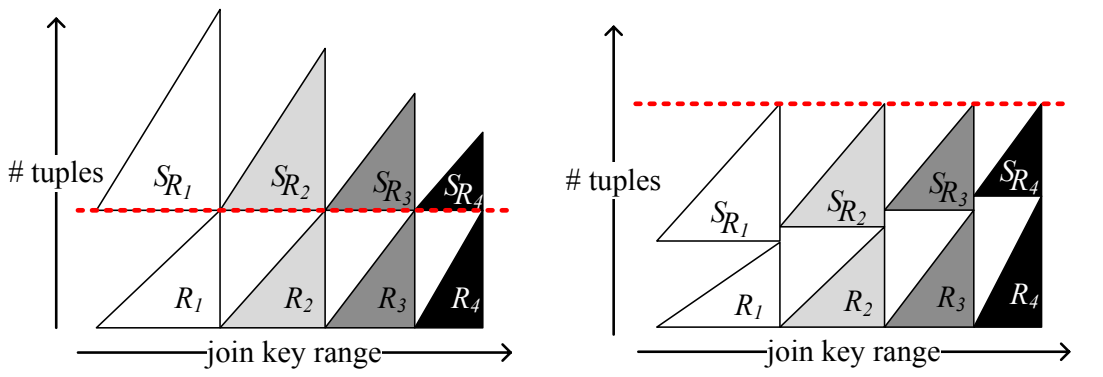

(a) Equi-height $R$ (left) vs. equi-cost *R-and-S* splitter partitioning (right)

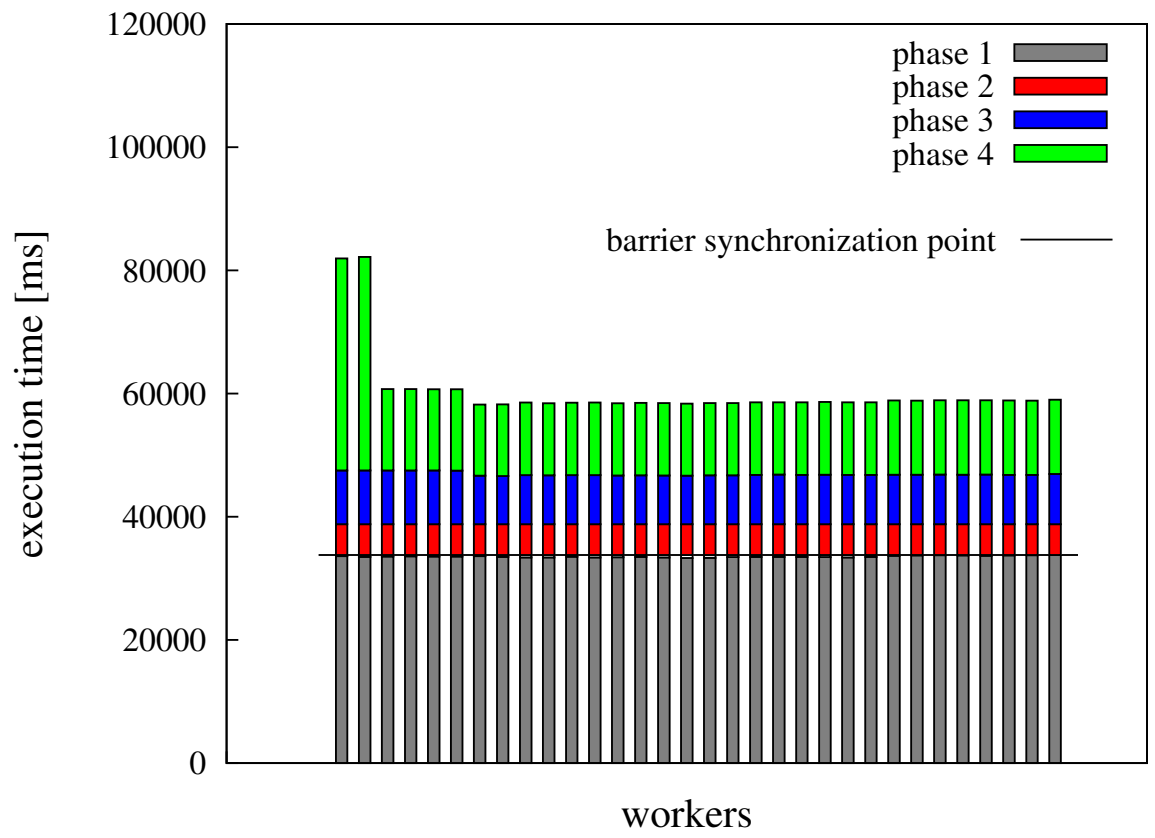

(b) Equi-height $R$ partitioning (multiplicity 4)

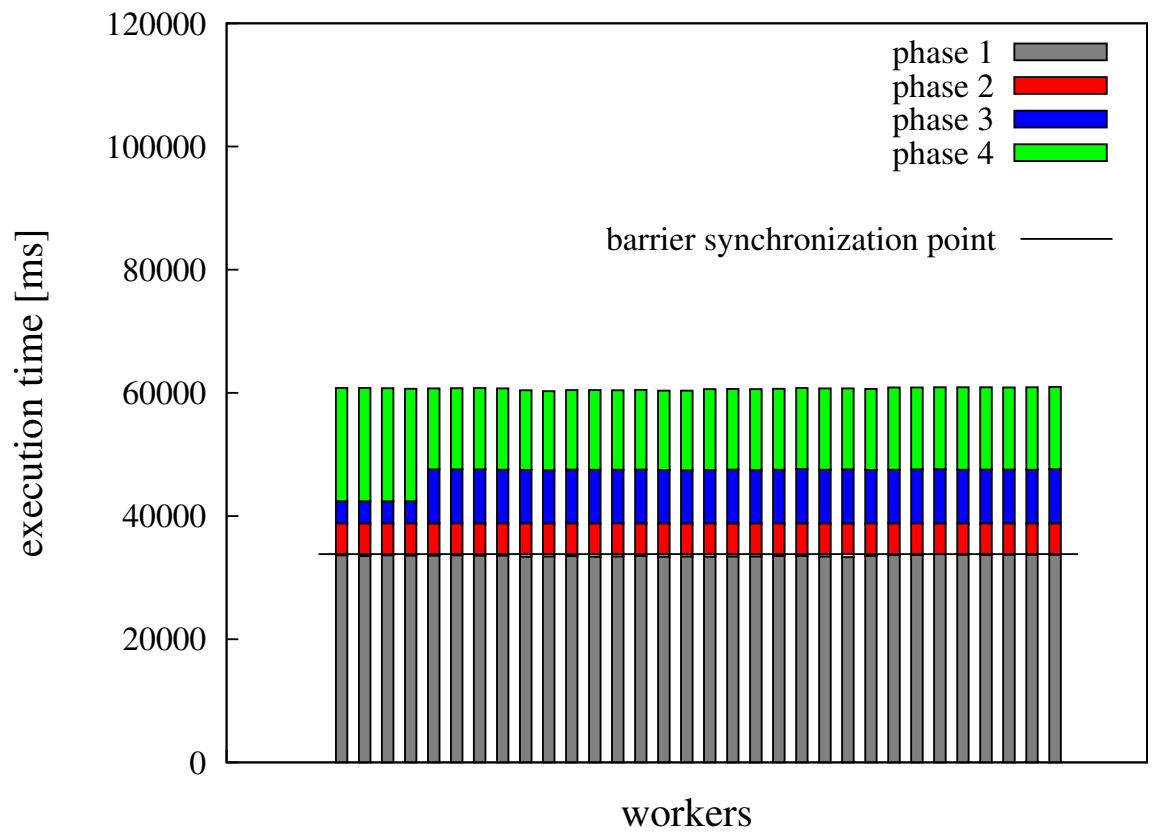

(c) Equi-cost *R-and-S* splitter partitioning (multiplicity 4)

Figure 16: Balancing splitters

cores, and smaller memory bandwidth per core sort-merge join is likely to outperform hash join on upcoming chip multiprocessors. Blanas et al. [1, 2] presented even better performance results for their parallel hash join variants. We compare the sort-based MPSM to their best-performing variant which we called Wisconsin hash join here and thereby pick the competition between sort-merge join and hash join up once again [12]. As a second contender we chose Vectorwise [15] that builds on the pioneering radix join work of MonetDB [4] in addition to vector-based processing of X100.

He et al. [14] develop parallel nested-loop, sort-merge, and hash joins on GPUs. The algorithms take advantage of massive thread parallelism, fast inter-processor communication through local memory, and histograms-based radix partitioning. We adapted the histogram approach for synchronization-free partitioning of MPSM's private input. For sorting in MPSM we developed our own Radix/IntroSort. In the future however, wider SIMD registers will allow to explore bitonic SIMD sorting [6].

MPSM does not produce completely sorted output. However, each worker's partition is subdivided into sorted runs. This interesting physical property might be exploited in further operations [7]. Note that the algorithms we compare MPSM to do not exhibit any interesting physical property in their output and we did **not** exploit any possibly pre-existing sorting in our comparative performance experiments.

Our disk-based D-MPSM was partly inspired by G-join [13] which also operates on sorted runs instead of hash partitions [12]. However, G-join lacks the parallelism which is in the focus of this paper.

# 7. CONCLUSIONS AND FUTURE WORK

The two dominating hardware trends are increasing RAM sizes and ever more (soon hundreds of) cores. Both facilitate the development of main-memory databases that are essential to propel the operational/real-time business intelligence applications. To minimize query response times (in our main memory database system HyPer) we devised a massively parallel algorithm for the most important query processing operator, the equi-join. MPSM merge joins in parallel sorted runs, which themselves were sorted by parallel threads. The performance analysis revealed that MPSM can effectively join very large main memory data of billions of tuples as it scales almost linearly with the number of cores. The scalable performance of MPSM is due to carefully exploiting the NUMA characteristics of the modern high-capacity servers. We avoided fine-grained synchronization and random access to remote NUMA memory partitions. The linear scalability in the number of cores promises MPSM to scale even beyond our tested 32 core, 1TB server – which is currently the top of the line main memory server but will soon be surpasseed by the next generations of servers with several TB capacity and hundreds of cores.

In future work we will develop the algorithmic details of MPSM for other join variants, e.g., outer, semi, and non-equi joins. Also, we are working on exploiting the "rough" sort order that MPSM inherently generates due to its range-partitioned run processing. This allows to optimize subsequent query plan operations analogously to traditional merge joins. In this paper we concentrated on the response time optimal range partitioned in-core variant of MPSM. In a follow-up paper we will also analyze in detail the memory constrained disk-based processing of D-MPSM that, due to space limitations, we could only sketch here. This variant is particularly promising for large batch query processing tasks that take place in parallel with transactions and real-time BI analytics.